\documentclass[prb,aps,twocolumn,citeautoscript,groupedaddress,noeprint,nofootinbib,longbibliography,floatfix]{revtex4-2}

\usepackage{amsmath}
\usepackage{amsthm}
\usepackage{comment}
\usepackage{amssymb}    % includes amsfonts, adds more symbols

\usepackage{bm}
\usepackage{graphicx}
\usepackage[dvipsnames]{xcolor}
\usepackage{tikz}
\usetikzlibrary{decorations.pathmorphing,arrows.meta,positioning,calc}
\usepackage{subfigure}
\usepackage{enumitem}

\newtheorem{lemma}{Lemma}
\newtheorem{theorem}{Theorem}

\usepackage{braket}
\usepackage{nicefrac}
\usepackage{placeins}
\usepackage{listings}
\usepackage{color}
\definecolor{myblue}{rgb}{0,0,1}
\definecolor{myred}{HTML}{E63946}
\definecolor{myblue}{HTML}{4361EE}
\definecolor{mygreen}{HTML}{4CAF50}
\usepackage[breaklinks=true,colorlinks=true,linkcolor=myblue,urlcolor=myblue,citecolor=myblue]{hyperref}

\usepackage[ruled,vlined]{algorithm2e}

\begin{document}
\title{CoS++: Towards More General and Explicit Implementations for Sampling High-Order Feynman Diagrammatic Series}

\author{Boyuan Shi}
\thanks{boyuanshi0607@gmail.com}

\affiliation{The Blackett Laboratory, Imperial College London, London SW7 2AZ, United Kingdom}

\begin{abstract}
We extend combinatorial summation (CoS) of connected Feynman diagrams to Hubbard models with flavor-dependent propagators and counterterm shifts. The construction combines an interaction-occupancy mask with labels for the open fermion cycle, allowing equivalent endpoint histories to be accumulated while retaining the flavor and sublattice information needed by subsequent contractions. In the flavor-dependent version, denoted symmetry-broken CoS (SB-CoS), normal heads carry the cycle flavor and counterterm heads inherit the flavor at their interaction partner. We prove the correctness of the resulting recurrence and introduce terminal readouts that return all nonempty interaction-subset values from a shared circuit, as required by a many-configuration Markov-chain sampler. For three flavors, a second formulation uses determinant polynomials and a local interaction tensor, with connected coefficients extracted by subset recursion or Fourier interpolation and a support sieve. Applications to the two-spin honeycomb and three-flavor square-lattice Hubbard models illustrate the treatment of sublattice structure and non-uniform reference shifts.
\end{abstract}

\maketitle

\section{Introduction}
At perturbative order $n$, a many-body calculation contains factorially many Wick contractions, whose large contributions of opposite sign can leave a much smaller connected sum. Diagrammatic Monte Carlo samples their spacetime integrals, including directly in the thermodynamic limit, and benefits from performing as much of the contraction sum as possible before sampling. Determinant identities and combinatorial recurrences achieve this by sharing the products that occur in many diagrams.

Hubbard models with enlarged flavor symmetry arise in alkaline-earth atomic gases \cite{Mott_SU_N,Gorshkov2010TwoOrbitalSUN} and in descriptions of moir\'{e} systems with approximate spin-valley symmetry \cite{SU_4_moire,YouVishwanath2019}. Their magnetic phases include spin liquids and valence-cluster states \cite{SU_N_spin_liquids,Ground_States_SU_N_Heisenberg,SU_N_phase_diagrams_AFM}. The larger local Hilbert space increases the cost of exact diagonalization \cite{SU_N_ED}, while determinant quantum Monte Carlo calculations face finite-size and sign limitations \cite{SU_N_DQMC}. In a diagrammatic expansion, the additional flavors enter through the allowed color assignments of the fermion loops.

For the two-spin Hubbard model, connected determinant methods start from a product of two determinants containing all contractions at fixed interaction vertices \cite{Rossi_CDet,sign_problem_despite_polynomial_complexity}. Fast principal-minor algorithms generate the subset values needed by the CDet recursion, which subtracts disconnected products \cite{fpm}. More general flavor structure introduces sums over local assignments; when propagators or counterterms depend on flavor, the color sum of a fermion loop also depends on its path.

Kozik's combinatorial summation (CoS) method represents the connected sum by a directed acyclic graph (DAG) \cite{CoS}. Its determinant construction draws on the Mahajan--Vinay dynamic program over ordered closed walks, or clow sequences, in which non-cycle-cover sequences cancel in sign-reversing pairs \cite{Mahajan1997Determinant}. CoS incorporates connectivity into the state and the allowed cycle openings, so the resulting circuit evaluates the connected sector directly.

This paper develops the state representations and readouts that CoS requires for contact interactions, sublattice sums, and flavor-dependent shifted expansions. We concentrate on evaluating the connected integrand and on passing its values to a Monte Carlo sampler; the numerical examples at selected state points illustrate the constructions.

For contact interactions, we replace the visited subset of $2n$ endpoints by occupancies
\[
m_i\in\{0,1,2\},
\]
that count the visited endpoints of interaction $i$. The mask has $3^n$ possible values, compared with $4^n$ possible endpoint subsets. On a bipartite lattice, a singly occupied interaction also records the sublattice of its visited endpoint; this label is dropped once the partner has been visited, while the head and current endpoint keep their own sublattice labels.

For non-uniform flavors, SB-CoS stores the color of each open cycle together with its normal or shift head type. Continuation preserves the cycle color, while a shift head inherits the color at its interaction partner, as required by the one-body counterterm. Summing these labeled paths replaces the uniform closed-loop factor $N_f$.

We attach terminal readouts to intermediate DAG layers to obtain connected values for all $2^N-1$ nonempty subsets of an $N$-vertex envelope. These values supply the heat-bath weights of a many-configuration Markov-chain sampler \cite{MCMCMC}. The construction retains the original interaction labels and prunes the graph with respect to all requested outputs.

For three flavors, we also formulate the sum as a polynomial with a local tensor for interactions and partner-attached shifts. Connected coefficients follow from subset recursion or, for a single final coefficient, Fourier interpolation combined with inclusion--exclusion over supports. This provides an algebraic check of the SB-CoS rules.

The Results section presents the constructions and numerical illustrations. The Methods section specifies the shifted action, gives the recurrences, and proves the typed-head construction.

\section{Results}
\subsection{Connected-sum recurrences}
At fixed interaction vertices, the two-spin Hubbard contraction sum is a product of two determinants, one per spin. In the flavor-uniform $\mathrm{SU}(N_f)$ formulation, contractions are represented on $2n$ endpoints with a factor $N_f$ for each closed fermion loop. Connectivity is determined by the union of the fermion edges and the interaction lines \cite{CoS}.

CoS-1 and CoS-2 denote the two constructions of Ref.~\cite{CoS}. CoS-1 retains a partition describing connectivity; CoS-2 orders the cycle openings through interaction partners and is the starting point here. We use SB-CoS for its flavor-dependent, typed-head extension. Table~\ref{tab:cos_terms} collects the terminology.

\begin{table*}
\caption{Terminology for the CoS variants used in this paper.}
\label{tab:cos_terms}
\begin{ruledtabular}
\begin{tabular}{p{0.18\textwidth}p{0.73\textwidth}}
Name & Meaning \\
CoS & The combinatorial summation framework of Ref.~\cite{CoS}, based on a Mahajan--Vinay type determinant DAG and a connectedness constraint on cycle openings. \\
CoS-1 & The component-partition construction of Ref.~\cite{CoS}. The algorithms developed here use CoS-2. \\
CoS-2 & The cycle-ordering construction of Ref.~\cite{CoS}.  Our uniform algorithms are CoS-2 variants with additional quotienting by contact-interaction lines. \\
Uniform CoS & The flavor-symmetric case in which every propagator block is identical and a closed loop contributes a single factor $N_f$. \\
SB-CoS & The symmetry-broken extension introduced here.  The state carries flavor labels and normal/shift head types, so flavor-dependent propagators, chemical potentials, and shifts can be treated explicitly. \\
CoS-GW & A topologically constrained variant mentioned in Ref.~\cite{CoS}; it is not used in the numerical calculations below. \\
All-subset readout & A terminal multi-output interface added to a CoS DAG so that one envelope evaluation returns connected values for all nonempty vertex subsets. \\
\end{tabular}
\end{ruledtabular}
\end{table*}

A DAG state records the head and endpoint of the open fermion cycle together with the visited interaction data. Continuation extends that cycle, while a close-and-open transition closes it and starts the next cycle at a canonically selected interaction partner, attaching the new cycle to the visited component and fixing its place in the traversal.

A determinant is a signed sum over cycle covers. For example, $(1)(2)(3,4)$ contributes $A_{11}A_{22}A_{34}A_{43}$ with sign $(-1)^{4-3}$. In the Mahajan--Vinay dynamic program \cite{Mahajan1997Determinant}, a state $(\ell,h,e)$ records the number of used edges, the current head, and the current endpoint. Its transitions continue the walk,
\begin{equation}
(\ell,h,e)\rightarrow(\ell+1,h,e'),
\qquad \text{weight } A_{ee'},
\end{equation}
or close the current walk and start a new one,
\begin{equation}
(\ell,h,e)\rightarrow(\ell+1,h',h'),
\qquad \text{weight } -A_{eh}.
\end{equation}
At the last layer the remaining open walk is also closed and accumulated into the sink.

The intermediate objects are \emph{clow sequences}, ordered lists of closed walks whose total length equals the matrix size. The head is the smallest vertex of a clow and occurs once; other vertices may repeat. Successive heads increase. Sequences that are not cycle covers cancel in sign-reversing pairs, leaving the determinant.

In the CoS-2 construction used below, the state additionally records the visited endpoint set $\mathcal R$. A continuation visits an unused endpoint, whereas a close-and-open step chooses an unused endpoint whose interaction partner lies in $\mathcal R$. After a union of complete cycles has been visited, connectedness guarantees a matching edge to its unvisited complement; a fixed ordering of the interactions selects one such edge and hence the next head. The proof of Algorithm~\ref{alg:cos_typed_heads} establishes the resulting traversal and its weights. Before contact-interaction symmetry is used, the $2^{2n}$ possible endpoint sets give an $\mathcal O(n^3 4^n)$ representation.

For contact interactions the two endpoints of the same interaction line are equivalent except for their role as head or current endpoint.  We therefore quotient the visited-set information by interaction lines.  The state records an interaction mask
\begin{equation}
\mathbf m=(m_1,\ldots,m_n),\qquad m_i\in\{0,1,2\},
\end{equation}
where $m_i$ counts how many endpoints of the $i$th interaction have been visited.  Continuation is allowed only if the target interaction has $m_i<2$ and the target endpoint is not the current head.  When a cycle is closed, the next head is chosen from a canonical open interaction with $m_i=1$, avoiding a repeated head.  This quotient reduces the dominant state count from $4^n$ to $3^n$ and gives the $\mathcal{O}(n^3 3^n)$ uniform contact-interaction algorithm used as the baseline in this work.

The quotient identifies endpoint histories with the same future transitions and preserves their multiplicities. Head and endpoint residues remain in the state because they determine the next matrix element. The resulting transition list can be compiled once and reused for different Monte Carlo configurations.

Bipartite lattices require one extra local label.  In the honeycomb calculations below, an interaction endpoint can lie on sublattice $A$ or $B$.  The quotient mask is therefore refined only when an interaction is singly occupied:
\begin{equation}
m_i\in\{0,(1,A),(1,B),2\}.
\end{equation}
If a continuation visits an unused interaction, both sublattice choices are generated.  If it visits the partner of a singly occupied interaction, the sublattice label is already fixed and is then discarded when the mask becomes $2$.  The same rule applies when a new cycle is opened.  More generally, for an interaction with $m$ internal labels the singly occupied state is split into $m$ possibilities, while the unvisited and fully visited states remain unique.

The stored sublattice fixes the partner endpoint of an onsite interaction. Head and current-endpoint labels also retain their sublattices for any subsequent propagator lookup.

With flavor-dependent propagators or shifts, SB-CoS attaches a color to each open cycle. Continuation preserves that color; opening a normal head sums over the allowed colors, and a shift head inherits its partner color. The frontier stores type and color while an interaction is unpaired. Algorithm~\ref{alg:cos_typed_heads} gives these transitions.

\subsection{All-subset readouts and many-configuration sampling}
The single-output circuit closes its final layer into one connected order-$n$ value. A many-configuration sampler instead uses an envelope
\begin{equation}
V=(X_{1},\ldots,X_{N})
\end{equation}
and assigns weights to retained subsets at each Monte Carlo step \cite{MCMCMC}. For allowed orders $\mathcal{U}$, define the signed flux
\begin{equation}
\Phi_{S}(V)=\lambda_{|S|} C_{|S|}(V_{S})
Q_{\bar S|S}(V_{\bar S}|V_{S}),
\qquad |S|\in\mathcal{U},
\end{equation}
where $C_{|S|}$ is the connected diagrammatic integrand of the retained subset and $Q_{\bar S|S}$ is the conditional density used to regenerate the complement. The corresponding local heat-bath normalization has the form
\begin{equation}
W(V)=\sum_{\substack{S\subseteq [N]\\ |S|\in\mathcal{U}}}
\frac{|\Phi_{S}(V)|}{\binom{N}{|S|}} .
\end{equation}
The quantities $C_{|S|}(V_S)$ share propagator entries and partial products. We compute them in a common circuit by adding rooted branches and terminal readouts to the CoS recurrence.

For the uniform $\mathrm{SU}(N_f)$ contact kernel, write the envelope size as $n=N$. The single-output state is
\begin{equation}
(h,e,\mathbf m),\qquad
\mathbf m=(m_{1},\ldots,m_{n}),\quad m_i\in\{0,1,2\},
\end{equation}
where $h$ and $e$ are the head and current endpoint residue of the open cycle and $m_i$ records whether the $i$th interaction has not been visited, has been visited once, or has been completed. To make every subset readout unique, we add a root label
\begin{equation}
r=\min S
\end{equation}
and propagate only through interaction labels $j\ge r$. The state becomes
\begin{equation}
(r,h,e,\mathbf m).
\end{equation}
Each nonempty subset belongs to the branch with $r=\min S$. Within a branch, continuation has the form
\begin{equation}
(r,h,e,\mathbf m)\longrightarrow
(r,h,j,\mathbf m+\mathbf e_j),
\qquad j\ge r,\quad m_j<2,
\end{equation}
with algebraic weight $(2-m_j)A_{ej}$ in the uniform contact case. Closing the current cycle and opening the canonical next head uses the same CoS rule as before: choose the first open interaction in the branch,
\begin{equation}
j=\min\{i\ge r:m_i=1\},
\end{equation}
set $m_j$ to $2$, and attach the factor $-N_f A_{eh}$. Other CoS variants only change the internal payload of the state and the algebraic labels on the edges. For $\mathrm{SU}(2)$ with unequal spin blocks one must keep the spin/block endpoint information; for bipartite lattices one keeps the sublattice label of singly occupied interactions; for non-uniform flavor shifts one keeps normal/shift head type and flavor labels for unpaired vertices. The root-label construction itself is independent of these choices.

For example, $\{1,3\}$ is read from branch $r=1$, and $\{2,3\}$ from branch $r=2$. The original interaction labels are retained so that each readout uses the corresponding principal submatrix of the envelope matrix.

For a subset $S$, the terminal interface $O_S$ is placed at layer
\begin{equation}
\ell(S)=2|S|-1
\end{equation}
and reads the states satisfying
\begin{equation}
r=\min S,\qquad
\mathbf m=2\mathbf 1_{S}.
\end{equation}
If $\mathcal{R}_{S}$ denotes this set of feeding states, the readout is
\begin{equation}
O_S =
\sum_{(r,h,e,\mathbf m)\in\mathcal{R}_{S}}
V_{\ell(S)}(r,h,e,\mathbf m)(-N_f A_{eh}) .
\end{equation}
There are $2^n-1$ output interfaces, each of which may sum several internal states. With $q_S=|\mathcal{R}_S|$, direct evaluation adds $\sum_S q_S$ final-close multiplications and $\sum_S\max(q_S-1,0)$ additions before factor sharing. These final-close operations are part of the circuit cost. The remaining overhead relative to a single-output graph comes from states that are kept only because they feed lower-order readouts, including the branches with $r>1$.

The output $O_S$ is terminal. Its feeding state can still propagate to larger-subset readouts through the original recurrence. Pruning retains every state that can reach at least one requested output, as described in the next subsection.
The resulting output vector supplies the connected values entering $W(V)$ in one shared evaluation.

\subsection{State pruning and circuit compression}
\label{sec:states_pruning}

We reduce the CoS circuit by canonicalizing states, removing branches that cannot reach a requested output, merging equal computations, and sharing repeated factor applications. These operations use the symbolic transition structure and its integer multiplicities.

Let
$\mathcal{G}=(\mathcal{V},\mathcal{E}),\qquad \mathcal{V}=\bigcup_{\ell=0}^{L}\mathcal{V}_{\ell}$,
be a layered DAG. A state $s\in\mathcal{V}_{\ell}$ carries a payload $x_s$, and an edge $e:s\rightarrow t$ carries an integer coefficient $c_e$ and an operator label $\lambda_e$. The label identifies a linear map $\mathsf L_{\lambda_e}$, giving the update
\begin{equation}
x_t\mathrel{+}=c_e\mathsf{L}_{\lambda_e}[x_s].
\label{eq:pruning_linear_update}
\end{equation}
For an ordinary scalar CoS graph,
$\mathsf{L}_{\lambda_e}[x]=f_{\lambda_e}x$,
where $f_{\lambda_e}$ identifies a Green's-function matrix element, a shift parameter, or a cycle-closing factor. Keeping the operator separate from its coefficient also permits vector and matrix payloads.

\paragraph{Canonical states.}

During construction, each state is stored under a canonical key $q(s)$. If $m$ unit-coefficient transitions have the same canonical child and operator label $\lambda$, they are accumulated as
\begin{equation}
\sum_{a=1}^{m}
\mathsf{L}_{\lambda}[x_s]
=
m\,\mathsf{L}_{\lambda}[x_s].
\label{eq:canonical_transition_multiplicity}
\end{equation}

For contact interactions, the two endpoints of interaction $i$ are represented by the occupancy
$m_i\in\{0,1,2\}$.
If an unvisited interaction is entered, its two endpoint choices give the same canonical child in the uniform case and are replaced by one edge with multiplicity $2$. More generally, a continuation into interaction $i$ has multiplicity
$2-m_i$.
The same principle is used to quotient identical spin blocks, to sort equivalent connected components into a canonical order, and to remove the absolute top/bottom endpoint label when only its residue modulo $n$ affects future matrix elements. For unequal spin blocks, flavor-dependent propagators, sublattice-dependent shifts, or normal/shift heads, the corresponding block, flavor, sublattice, or head-type label is retained because it changes the future algebraic operators. Canonicalization therefore removes only information that is provably irrelevant to every continuation.

Histories with the same key contribute to one layer value. This accumulation is part of the recurrence and precedes pruning of the completed DAG.

\paragraph{Local rejection tests.}

Necessary connectivity and ordering conditions can reject states before their outgoing edges are generated.

For CoS-1, the state records the connected components formed by the interaction lines between fermion loops. A closed component that has no unconsumed interaction endpoint through which it can be joined to the current component is permanently stranded and cannot contribute to a connected final diagram. Such a state is removed immediately. A second necessary condition compares the number $c(s)$ of current components with the number $r(s)$ of remaining interaction lines capable of joining them. Since at least $c(s)-1$ connections are required,
\begin{equation}
r(s)<c(s)-1
\quad\Longrightarrow\quad
s\ \text{is unproductive}.
\label{eq:connectivity_edge_bound}
\end{equation}
In the uniform $\mathrm{SU}(N_f)$ CoS-1 construction, an interaction whose two endpoints lie on the same fermion loop gives a self-edge in the loop-coloring graph. Its proper-coloring multiplicity is zero, so the corresponding state can also be discarded exactly.

Ordering restrictions provide another local reduction. Once an all-vertices or canonical-frontier ordering has been imposed, a state that cannot satisfy the remaining ordering constraints is excluded. Similarly, an edge known to contain an identically zero matrix element may be deleted, as in the explicitly zero-diagonal specialization. Such a deletion is allowed only when the zero follows from the model assumptions; accidental zeros in a particular random or Monte Carlo matrix are not used for structural pruning.

Early rejection avoids generating branches with no admissible nonzero contribution. The final recurrence still enforces the connectivity and ordering constraints, while algebraic zero tests can remove branches that remain connected to an output in the unweighted graph.

\paragraph{Reachability and state equivalence.}

After the structural DAG has been generated, we retain only states lying on a path from an initialized state to at least one requested output. Let $\mathcal{I}\subseteq\mathcal{V}_0$ be the initialized states. The forward-reachable sets are
\begin{align}
\mathcal{F}_0&=\mathcal{I},\\
\mathcal{F}_{\ell+1}
&=
\left\{
t\in\mathcal{V}_{\ell+1}:
\exists\,s\in\mathcal{F}_{\ell}
\text{ with }(s,t)\in\mathcal{E}
\right\}.
\label{eq:state_pruning_forward}
\end{align}
Let $\mathcal{R}_{\ell}$ contain the states in layer $\ell$ that feed a requested terminal interface. The backward-productive sets are
\begin{align}
\mathcal{B}_L&=\mathcal{R}_L,\\
\mathcal{B}_{\ell}
&=
\mathcal{R}_{\ell}
\cup
\left\{
s\in\mathcal{V}_{\ell}:
\exists\,t\in\mathcal{B}_{\ell+1}
\text{ with }(s,t)\in\mathcal{E}
\right\}.
\label{eq:state_pruning_backward}
\end{align}
The surviving states are
\begin{equation}
\mathcal{K}_{\ell}
=
\mathcal{F}_{\ell}\cap\mathcal{B}_{\ell}.
\label{eq:state_pruning_keep}
\end{equation}

For a single-output circuit, $\mathcal{R}_{\ell}$ is empty except at the final layer. For the all-subset circuit,
\begin{equation}
\mathcal{R}_{\ell}
=
\bigcup_{\substack{\varnothing\neq S\subseteq[N]\\
                   \ell(S)=\ell}}
\mathcal{R}_S,
\qquad
\ell(S)=2|S|-1.
\end{equation}
A state that is dead with respect to the full-order sink may therefore remain live because it feeds a lower-order output $O_S$. The scalar output $O_S$ itself is terminal and is not propagated further.

This deletion is exact because every output is a sum over directed paths. A state outside $\mathcal{F}_{\ell}$ has no initialized ancestor, while a state outside $\mathcal{B}_{\ell}$ has no requested descendant. It therefore belongs to no source-to-output path and contributes to no requested value.

The surviving graph is further quotiented by exact equivalence of its arithmetic circuits. We first coalesce parallel edges with the same source, destination, and operator label by adding their integer coefficients. Terms whose total coefficient vanishes are removed.

Future equivalence is computed in reverse topological order. After the child classes in layer $\ell+1$ have been determined, the complete outgoing signature of $s\in\mathcal{K}_{\ell}$ is
\begin{equation}
\Sigma^{+}_{\ell}(s)
=
\left(
\rho^{+}(s),
\operatorname{sort}
\left\{
\bigl(\mathsf{L},C,\kappa_s(\mathsf{L},C)\bigr):
\kappa_s(\mathsf{L},C)\neq0
\right\}
\right),
\label{eq:future_state_signature}
\end{equation}
where $C$ is a child equivalence class, $\mathsf{L}$ is an exact operator label, and
\begin{equation}
\kappa_s(\mathsf{L},C)
=
\sum_{\substack{e:s\rightarrow t\\
                 \mathsf{L}_{\lambda_e}=\mathsf{L},\ [t]=C}}
c_e .
\end{equation}
The boundary annotation $\rho^{+}(s)$ includes the complete direct-readout map, with a distinct identity for each subset output. Equality of the full signatures establishes future equivalence: incoming values are added in the merged state, which retains one copy of the common continuation. All outgoing operators, child classes, and multiplicities enter this comparison.

A complementary top-down reduction identifies states with the same past computation. Once parent classes have been fixed, define
\begin{equation}
\Sigma^{-}_{\ell}(s)
=
\left(
\rho^{-}(s),
\operatorname{sort}
\left\{
\bigl(P,\mathsf{L},\kappa_s^{-}(P,\mathsf{L})\bigr):
\kappa_s^{-}(P,\mathsf{L})\neq0
\right\}
\right),
\label{eq:past_state_signature}
\end{equation}
where
\begin{equation}
\kappa_s^{-}(P,\mathsf{L})
=
\sum_{\substack{e:p\rightarrow s\\
                 [p]=P,\ \mathsf{L}_{\lambda_e}=\mathsf{L}}}
c_e,
\end{equation}
and $\rho^{-}(s)$ preserves initialized-state information. States with equal past signatures carry the same payload for every admissible input. One copy of their incoming computation is retained, and their outgoing uses are combined with multiplicities.

Future and past merges can expose new parallel edges and cancellations, so the passes are alternated with edge coalescing and productivity pruning until the graph stabilizes.

\paragraph{Vector and matrix payloads.}

The preceding reductions are not restricted to scalar states. In the literal DAG, a state may explicitly store the current head $h$ and endpoint $e$. When spin blocks are identical, these labels can instead be moved into the payload. A tail-vector representation stores
$x_s(e)$,
while a more compressed representation stores
$X_s(h,e)$.
The structural state then contains only the block, occupancy, and component information. Continuation, cycle closing, and opening a new head become linear maps acting on the vector or matrix payload.

Typical operators include multiplication by the common Green's-function block, projection onto an allowed row or column, diagonal contraction when a cycle is closed, and insertion of the resulting scalar into the coordinate associated with a new head. For example, several continuation branches may share the intermediate product
$Y_s=X_s B$,
after which different child operators select different columns or projected components of $Y_s$. The matrix multiplication is performed once for the source state rather than once for every outgoing branch.

For vector and matrix payloads, signatures record the complete linear operator, including projections and component placement. Future equivalence requires equal maps to the outputs; past equivalence requires equal expressions for the payloads. Both identities must hold for all inputs satisfying the assumed block symmetries.

\paragraph{Sharing factor applications.}

State and edge reduction determines the arithmetic circuit, but the same circuit may still admit different numerical evaluation schedules. Consider all scalar edges in one layer carrying the same factor $f$. Their contribution can be written as
\begin{equation}
y_d
\mathrel{+}=
f\sum_s c^{(f)}_{sd}x_s.
\label{eq:factorized_layer_update}
\end{equation}
There are two equivalent evaluation orders. In a source-side schedule one computes
$z_s=f x_s$
once for each participating source and scatters $c^{(f)}_{sd}z_s$ to its destinations. In a destination-side schedule one first forms
\begin{equation}
u_d=\sum_s c^{(f)}_{sd}x_s
\end{equation}
and then computes $f u_d$ once for each participating destination.

Let $S_f$ and $D_f$ be the sets of source and destination states incident to edges with label $f$. Choosing the cheaper side gives
\begin{equation}
N_{\mathrm{mult}}(f)
=
\min\bigl(|S_f|,|D_f|\bigr)
\label{eq:min_side_factor_count}
\end{equation}
factor multiplications for that layer, rather than one multiplication per edge. Final-closing edges are grouped in the same way: all contributions carrying the same final factor are summed before that factor is multiplied.

Equation~\eqref{eq:min_side_factor_count} counts only multiplications by $f$; integer coefficients and the accompanying additions are counted separately. Matrix payloads admit the analogous sharing of $X_sB$ across several projections.

The reduced DAG and schedule are stored as state identifiers, operator labels, coefficients, and source/destination indices and reused at fixed order and model variant.

\subsection{Flavor-polynomial route for non-uniform shifts}
For three flavors, the contraction sum can be organized as a polynomial. Principal-minor coefficients collect the contractions of each flavor, and a local tensor specifies the interaction and counterterm at each occupied site. A formal logarithm extracts the connected part. We use this representation as an independent check of the typed-head rules and describe a Fourier evaluation for the final coefficient.

Let $a=0,1,2$ label the three flavors. For fixed interaction vertices $X_{1},\ldots,X_{n}$, define the flavor-resolved Green's-function matrix
\begin{equation}
A^{(a)}_{ij}=G_{a}(X_i,X_j).
\end{equation}
The reference propagator includes the flavor-dependent chemical potential and uses the equal-time convention $G_a(0^-)=-\rho_a$. The shift-close weight is $\sigma_a=\alpha_a/U$ for physical shift $\alpha_a$, so the uniform specialization is
\begin{equation}
\begin{gathered}
A^{(0)}=A^{(1)}=A^{(2)}=A,\\
\sigma_0=\sigma_1=\sigma_2=\sigma=\alpha/U.
\end{gathered}
\end{equation}
These are the same shift-close weights used in Algorithm~\ref{alg:cos_typed_heads}.

For each flavor we first introduce the principal-minor polynomial
\begin{equation}
\begin{aligned}
D_a(\mathbf z_a)
&=\det\!\left(I+A^{(a)}
\operatorname{diag}(z_{a,1},\ldots,z_{a,n})\right)\\
&=\sum_{S\subseteq[n]} d_a(S)\prod_{i\in S}z_{a,i}.
\end{aligned}
\end{equation}
Here $d_a(S)=\det A^{(a)}[S,S]$, with $d_a(\varnothing)=1$. The coefficients can be obtained by fast principal minors or binary Fourier interpolation. A Gray-code traversal changes one column at a time, allowing rank-one determinant updates when the intermediate matrices are nonsingular.

Let $m_a\in\{0,1\}$ indicate whether flavor $a$ occupies site $i$ in its determinant coefficient. The local tensor is
\begin{equation}
L_i(\mathbf m;x_i)=
\begin{cases}
1, & \mathbf m=(0,0,0),\\
2\sigma_a x_i, & \mathbf m=\mathbf e_a,\\
2x_i, & \mathbf m=\mathbf e_a+\mathbf e_b,\ a<b,\\
0, & \mathbf m=(1,1,1).
\end{cases}
\end{equation}
The factor $2$ counts the two endpoint assignments of a contact interaction. A one-flavor occupation pairs a retained fermion bilinear of color $a$ with its counterterm, giving weight $2\sigma_a$. Two distinct flavors form an ordinary interaction pair, while triple occupation is excluded. Thus the singleton entry implements the same partner-color rule as a shift head in SB-CoS.

Combining the determinant polynomials with the local tensor gives the full-moment generating polynomial
\begin{equation}
\begin{aligned}
H(\mathbf x)=
\sum_{S_0,S_1,S_2\subseteq[n]}
&d_0(S_0)d_1(S_1)d_2(S_2)\\
&\times
\prod_{i=1}^{n}
L_i(\chi_{0i},\chi_{1i},\chi_{2i};x_i),
\end{aligned}
\end{equation}
where $\chi_{ai}=\mathbf 1_{i\in S_a}$.
The coefficient $H(T)=[x_T]H(\mathbf x)$, with $x_T=\prod_{i\in T}x_i$, is the full moment on the retained interaction set $T$. The physical local insertion contains no shift--shift term. Some auxiliary endpoint formulations include a scalar completion with weight $q$; its subset polynomial is
\begin{equation}
Q(\mathbf x)=\prod_{i=1}^{n}(1+q x_i),
\qquad
F(\mathbf x)=[H(\mathbf x)Q(\mathbf x)]_{\mathrm{sf}},
\end{equation}
where $[\cdot]_{\mathrm{sf}}$ denotes reduction modulo $x_i^2=0$. In this algebra, $\log F=\log H+q\sum_i x_i$, so the completion changes only singleton coefficients. The raw typed circuit gives $q=\sum_a\sigma_a^2$, whereas an independent-color shift completion gives $q=(\sum_a\sigma_a)^2$. Subtracting the corresponding $q$ from each singleton restores the physical convention. The polynomial $H$ already has that convention.

For all-subset output, the physical connected coefficients follow from the CDet recursion on $H$:
\begin{equation}
C(T)=H(T)-\sum_{\substack{S\subsetneq T\\ r(T)\in S}}
C(S)H(T\setminus S),
\qquad r(T)=\min T .
\end{equation}
For final-only evaluation there is a more direct form. Define $H_S(t)$ by setting $x_i=t$ for $i\in S$ and $x_i=0$ otherwise. The squarefree connected coefficient is recovered by the support sieve
\begin{equation}
[x_1\cdots x_n]\log H(\mathbf x)
=\sum_{S\subseteq[n]}(-1)^{n-|S|}
[t^n]\log H_S(t).
\end{equation}
The polynomial $H_S(t)$ has degree at most $|S|$, so $|S|+1$ Fourier samples determine its coefficients. The formal logarithm is then computed from those coefficients through degree $n$. Inclusion--exclusion removes degree-$n$ terms whose support omits an interaction label; the surviving term is squarefree.

For generic three-flavor parameters, the $2\times2\times2$ local tensor has a rank-two decomposition over the complex numbers. Each Fourier sample then becomes a sum of products of three weighted determinants. Rank-one determinant updates accelerate successive evaluations when the intermediate matrices are nonsingular. Singular matrices are handled by recomputing the determinants, and degenerate local tensors by direct contraction.

The subset recursion and Fourier construction evaluate the same connected polynomial, providing an independent comparison with SB-CoS for non-uniform matrices and shifts. In the uniform specialization, this also agrees with the loop-factor formulation using the true-diagonal shift $\sigma/N_f$, with each route's singleton correction applied.

\subsection{Numerical examples}
We consider the two-spin Hubbard model on a honeycomb lattice, which uses the sublattice-resolved interaction masks described above, and the three-flavor model on a square lattice, evaluated with uniform CoS and SB-CoS. The observables are the logarithmic partition function, double occupancy, and thermodynamic derivatives. We display order-by-order partial sums and Pad\'{e} continuations to assess their dependence on truncation and on the reference shifts.

\begin{figure*}[t]
    \centering
    \includegraphics[width=\textwidth]{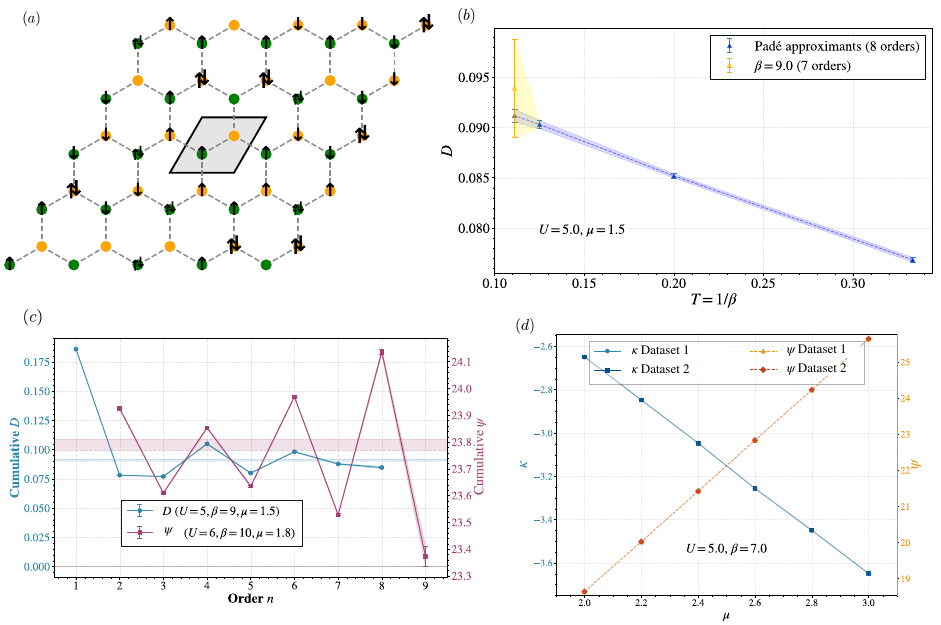}
    \caption{Two-spin Hubbard model on a honeycomb lattice with $32\times32$ unit cells and 2048 sites. (a) Lattice geometry. (b) Pad\'{e} estimates of double occupancy at $\beta = 3, 5, 8, 9$, including the seventh- and eighth-order comparison at $\beta=9$. (c) Cumulative double occupancy $D$ and logarithmic partition density $\psi=\ln\mathcal Z/N_s$ at the indicated parameters. (d) Logarithmic partition density $\psi$ and grand-energy density $\kappa=-\partial_\beta\psi$ at $U/t=5$, $\beta=7$. Datasets 1 and 2 use different auxiliary shifts. The blue lines join the grand-energy estimates; the orange lines join the logarithmic partition estimates.}
    \label{fig:SU_2_honeycomb}
\end{figure*}

For the honeycomb model we report the grand canonical partition function
\begin{equation}
\mathcal{Z}(\beta,\mu)=\mathrm{Tr}\,e^{-\beta(H-\mu N)}
\end{equation}
and the site-averaged double occupancy
\begin{equation}
D=\frac{1}{N_s}\sum_i\langle n_{i\uparrow}n_{i\downarrow}\rangle .
\end{equation}
Here $N_s$ is the number of lattice sites. The figures use the densities $\psi=\ln\mathcal Z/N_s$ and $\kappa=\langle K\rangle/N_s$, where $K=H-\mu N$. In terms of $\ln\mathcal Z$, the thermodynamic identities are
\begin{equation}
\partial_\mu \ln\mathcal{Z}=\beta\langle N\rangle,\qquad
\partial_\beta\ln\mathcal{Z}=-\langle K\rangle.
\end{equation}
Differentiating the first identity with respect to $\beta$ at fixed $\mu$ gives
\begin{equation}
\frac{\partial^2\ln\mathcal{Z}}{\partial\beta\,\partial\mu}
=\langle N\rangle+\beta\frac{\partial\langle N\rangle}{\partial\beta}.
\end{equation}
It follows that $\partial_\mu\kappa=-n-\beta\partial_\beta n$, with total density $n=\langle N\rangle/N_s$. On a density plateau that depends weakly on temperature, the slope approaches $-n$. Between $\mu/t=2.0$ and $3.0$, the endpoint-to-endpoint slopes of the two grand-energy curves in Fig.~\ref{fig:SU_2_honeycomb}(d) are approximately $-1.00$; those of $\psi$ are $6.99$ and $7.00$ for Datasets 1 and 2, respectively. These values are consistent with $n\approx1$ on the half-filled plateau and with $\partial_\mu\psi=\beta n$ at $\beta=7$.

Panels (b) and (c) use Pad\'{e} continuation of the shifted interaction series. For coefficients $\{a_i\}$ with uncertainties $\{\delta a_i\}$, we use several numerator and denominator degrees. Under the approximation of independent coefficient errors, the propagated uncertainty is
\begin{equation}
\delta I =
\sqrt{\sum_i
\left|
\frac{\partial\,\mathrm{Pad\acute e}(\{a_i\};m,n)}
{\partial a_i}
\right|^2\delta a_i^2}.
\end{equation}
The band is the envelope of $I\pm\delta I$ over the retained approximants. It combines the propagated coefficient errors with the spread of these continuations; it does not bound the omitted orders. At $\beta=9$, adding the eighth-order coefficient narrows the displayed band. Panel (c) gives the partial sums of $D$ at $U=5.0$, $\mu=1.5$, $\beta=9.0$ and of $\ln\mathcal{Z}$ at $U=6.0$, $\mu=1.8$, $\beta=10.0$.

Panel (d) compares two reference shifts at matched physical parameters. The two grand-energy curves follow the same trend, but at $\mu/t=2.8$ their estimates, $-3.450830\pm0.000386$ and $-3.448359\pm0.000398$, differ by more than the displayed Pad\'{e} intervals (a relative difference of $7\times10^{-4}$). Since the exact result is independent of the reference shift, this difference reveals residual dependence on truncation and continuation at the available orders.

\begin{figure*}[t]
    \centering
    \includegraphics[width=\textwidth]{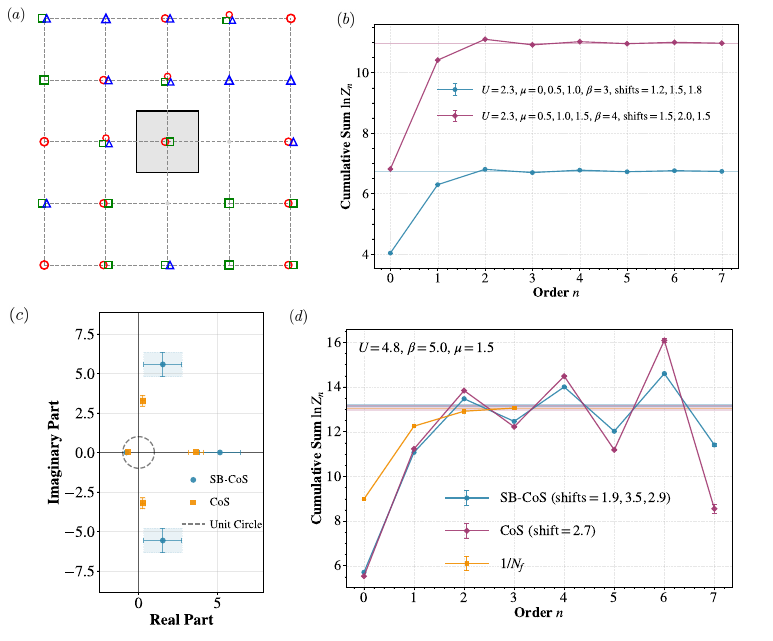}
    \caption{Three-flavor Hubbard model on a $32\times32$ square lattice. (a) Lattice and flavor schematic. (b) Cumulative logarithmic partition values for the two non-uniform parameter sets shown in the legend. (c) Poles of $[2/4]$ Pad\'{e} approximants constructed from the coefficients through sixth order for the series in panel (d). (d) Uniform CoS and SB-CoS partial sums and continuations, together with the $1/N_f$ result. Logarithmic quantities use the per-site normalization. The successive $1/N_f$ points label stages of the large-flavor expansion, independently of the CoS interaction order.}
    \label{fig:SU_3}
\end{figure*}

Figure~\ref{fig:SU_3}(d) compares uniform CoS and SB-CoS at matched physical parameters with different reference shifts. Their Pad\'{e} bands overlap. The negative-real poles of the rational approximants in panel (c) are consistent with a convergence radius below unity and with the alternating partial sums in panel (d).

Panel (d) also includes the $1/N_f$ expansion through $\mathcal{O}(1/N_f^2)$ from Ref.~\cite{MY_1_OVER_NF_PAPER}, whose final value lies slightly below the CoS and SB-CoS continuations. In panel (b), the two sets of flavor-dependent chemical potentials and shifts describe different physical state points. SB-CoS evaluates both cases with separate color labels, and in both cases the partial sums change little beyond second order.

\section{Discussion}
The constructions developed here extend CoS-2 to flavor-dependent propagators and counterterms while retaining a direct sum over connected contractions. Contact symmetry reduces the endpoint record to $3^n$ occupancy patterns in the uniform case; sublattice and typed-head labels retain the additional information needed for the honeycomb and non-uniform calculations. The path construction and weighted quotient establish that these reductions preserve the connected sum.

The all-subset circuit returns $2^N-1$ connected values from shared intermediate computations. Together with pruning relative to the requested outputs, it gives CoS the interface needed by many-configuration sampling. The three-flavor polynomial construction supplies a second representation of the same shifted coefficients, evaluated by subset recursion or a final-only Fourier and support-sieve calculation.

The numerical examples demonstrate the use of these evaluators at non-uniform reference shifts. In the honeycomb calculation, the eighth-order coefficient narrows the displayed continuation band at $\beta=9$; for the three-flavor square-lattice model, the CoS and SB-CoS continuations have overlapping bands, and SB-CoS also treats unequal physical chemical potentials. These results motivate extending the polynomial contraction to more flavors and using the shared subset readouts in production many-configuration simulations.

\section{Methods}
\subsection{Models and diagrammatic expansions}
\label{sec:shift_conventions}
We add a one-body shift to the free action and subtract it through the counterterm. The free propagator is evaluated at the shifted chemical potential. Multiplying the interaction and compensating counterterm by $\xi$ defines the expansion; setting $\xi=1$ recovers the physical action.

Writing $n_{i,\sigma}=\bar\psi_{i,\sigma}\psi_{i,\sigma}$, the imaginary-time action is
\begin{equation}
\begin{aligned}
S(\xi)={}&\int_0^\beta d\tau\,
\Bigg\{\sum_{i,j,\sigma}\bar\psi_{i,\sigma}
\Big[\delta_{ij}(\partial_\tau-\mu_\sigma+h_{i,\sigma})
+h_{ij}\Big]\psi_{j,\sigma}\\
&\quad+\xi\left[
\frac U2\sum_{i,\sigma,\sigma'}n_{i,\sigma}n_{i,\sigma'}
-\sum_{i,\sigma}h_{i,\sigma}n_{i,\sigma}\right]\Bigg\}.
\end{aligned}
\end{equation}
All fields in the integrand are evaluated at $\tau$. Here $h_{ij}$ denotes hopping, $\mu_\sigma$ the flavor chemical potential, $U$ the onsite coupling, and $h_{i,\sigma}$ the auxiliary one-body shift. The formal flavor-pair sum uses Grassmann densities, for which the same-flavor contact term vanishes. Its operator counterpart is the normal-ordered interaction.

For a physical observable $F$ we expand in powers of $\xi$,
\begin{equation}
F=\sum_n f_n\xi^n .
\end{equation}
The coefficients are obtained from the connected expansion appropriate to $F$. For the two-spin Hubbard model, CDet and CoS provide alternative evaluations of the connected integrand.

For the honeycomb reference we use chemical-potential and staggered-field shifts. With nearest-neighbor hopping, the free Hamiltonian is
\begin{equation}
H_{0}=\sum_{\bm k}\sum_{\sigma=\uparrow,\downarrow}
\bm{c}^{\dagger}_{\sigma}(\bm{k})\,\mathcal{H}_{\sigma}(\bm{k})\,\bm{c}_{\sigma}(\bm{k}),
\end{equation}
with
\begin{equation}
\mathcal{H}_{\sigma}(\bm{k})=
\begin{bmatrix}
\Delta_{\sigma}-(\mu-\alpha_{\sigma}) & \epsilon(\bm{k}) \\
\epsilon^{\ast}(\bm{k}) & -\Delta_{\sigma}-(\mu-\alpha_{\sigma})
\end{bmatrix},
\end{equation}
where $\epsilon(\bm{k})=-(1+e^{i\bm{k}\cdot \bm{b}_{1}}+e^{i\bm{k}\cdot \bm{b}_{2}})$, with hopping set to one. The displayed honeycomb reference has a common chemical potential $\mu$. The shifts on its two sublattices are $h_{A,\sigma}=\alpha_\sigma+\Delta_\sigma$ and $h_{B,\sigma}=\alpha_\sigma-\Delta_\sigma$. On the square lattice, $\mathcal{H}_{\sigma}(\boldsymbol{k})=-2[\cos(2\pi k_{x}/N_{1})+\cos(2\pi k_{y}/N_{2})]-\mu_\sigma+\alpha_{\sigma}$.

With the contraction convention $g_\sigma(0^-)=-\langle n_\sigma\rangle_0$, the two-spin counterterms enter the true diagonal as
\begin{equation}
\begin{aligned}
g^{AA(BB)}_\sigma(0^-;\bm R_i,\bm R_i)
&\ \mapsto\ g^{AA(BB)}_\sigma(0^-;\bm R_i,\bm R_i)\\
&\qquad+\frac{\alpha_{\bar\sigma}\pm\Delta_{\bar\sigma}}{U},
\end{aligned}
\end{equation}
where $\bar\sigma$ is the opposite spin. The spin reversal follows from
\begin{equation}
\begin{aligned}
Un_\uparrow n_\downarrow-h_\uparrow n_\uparrow-h_\downarrow n_\downarrow
={}&U(n_\uparrow-h_\downarrow/U)(n_\downarrow-h_\uparrow/U)\\
&-h_\uparrow h_\downarrow/U.
\end{aligned}
\end{equation}
The scalar term corrects the first-order logarithmic coefficient.

In SB-CoS the physical shift $\alpha_a$ has shift-close weight $\sigma_a=\alpha_a/U$. A uniform loop-factor calculation uses the true-diagonal weight $\alpha/(N_fU)$, since closure supplies $N_f$. Propagators between distinct endpoints at the same residue retain their normal values.

For the two-spin construction, define the order-$m$ integrated moment with its coupling factor stripped off by
\begin{equation}
\mathcal A^{(m)}=\frac{1}{m!}\!
\int_{X_{1},\ldots,X_{m}}
\sum_{a_{1},\ldots,a_{m}}
\prod_{\sigma}
\det \mathbf{G}^{(\sigma)}[\{X_{i},a_{i}\}],
\end{equation}
where the $m\times m$ matrix $\mathbf{G}^{(\sigma)}$ has entries
\begin{equation}
\big[\mathbf{G}^{(\sigma)}\big]_{ij}
= g^{a_{i}a_{j}}_{\sigma}(X_{i},X_{j}), 
\qquad 1\leq i,j \leq m.
\end{equation}
Here $g$ includes the diagonal counterterms just specified. Multiplication by $(-\xi U)^m$ gives the contribution of the shifted product vertex; the scalar correction above restores the physical action.

For the vacuum diagrams considered here, let $a(V)$ be the full contraction sum at fixed vertices, with $a(\varnothing)=1$, and let $c(V)$ be its connected part. Choose an anchor $r(V)=\min V$. Every disconnected diagram has one component containing the anchor, which gives the recursion
\begin{equation}
c(V)=a(V)-\sum_{\substack{S\subsetneq V\\r(V)\in S}}
c(S)\,a(V\setminus S).
\label{eq:anchored_connected_recursion}
\end{equation}
The sum runs over proper subsets containing the anchor and is evaluated in increasing subset size.

The sublattice assignments can be summed at fixed unit-cell and time coordinates,
\begin{equation}
a(W)=\sum_{s\in\{\text{sublattice conf.}\}}a(W;s), 
\end{equation}
\begin{equation}
c(W)=\sum_{s}c(W;s),
\end{equation}
using a three-branch principal-minor construction with $\mathcal{O}(3^n)$ scaling. Alternatively, sublattice labels can be sampled along with spacetime coordinates, with the corresponding proposal probabilities included in the estimator.

For CoS and SB-CoS, the recurrences below construct the connected contraction sum with the retained flavor and sublattice labels.

\subsection{Fast generation of $3^n$ minors}
The restricted principal minors are generated by a depth-first Schur-complement traversal. At depth $d\in[0,n-1]$, the active matrix is $M^{(d)}\in\mathbb{R}^{2(n-d)\times2(n-d)}$. The three branches omit the current interaction, select its first sublattice entry, or select its second entry. After selecting an entry, the unused alternative is discarded before the next interaction is processed. When a pivot $p$ is too small, choose a nonsingular replacement $\widetilde p$ and record the diagonal increment
\begin{equation}
\delta=\widetilde p-p.
\end{equation}
For a zero pivot, one may set $\widetilde p=C_{\mathrm{shift}}>0$ using the matrix scale, in which case $\delta=C_{\mathrm{shift}}$. Multilinearity gives $\det(M+\delta e_ie_i^{\mathsf T})=\det M+\delta\det M_{\setminus i}$, so the added contribution is subtracted after traversal. For depth $d$, choice $c\in\{1,2\}$, prefix $b$, and $k=3^{n-d-1}$, the update is
\begin{align}
&\mathrm{Minor}[(3b+c)k+s]\nonumber\\
&\qquad\mathrel{-}=\delta\,\mathrm{Minor}[(3b)k+s].
\label{eq:pivot_correction}
\end{align}
Here $s\in[0,k)$ indexes the remaining choices. The branch prefactor is already included in both stored minors. Corrections are applied deepest first, using the recorded $\delta$ at each modified pivot, so the same update also handles small nonzero pivots.

The determinant-recursion calculations use double-double precision.

\subsection{Three combinatorial algorithms}
Algorithms~\ref{alg:cos_sublattice}, \ref{alg:cos_dedup}, and \ref{alg:cos_typed_heads} give the two-spin sublattice sum, the flavor-uniform $\mathrm{SU}(N_f)$ sublattice sum, and the typed-head construction, respectively. Each layer consumes one endpoint. A state retains the current head, endpoint, and interaction record. Continuation multiplies a propagator; closure supplies the cycle sign and factor before opening the next canonical head. The local record distinguishes the sublattice or flavor choices required by each model.

\begin{figure*}
\begin{algorithm}[H]
\DontPrintSemicolon
\caption{Connected Sum for $\text{SU}(2)$ Bipartite Model}
\label{alg:cos_sublattice}
\KwIn{$n$ (interaction pairs)}
\KwOut{Sum $\mathcal{S}$}
$N \leftarrow 2n$; $\mathcal{I} \leftarrow \{(i, i+n) : i \in [n]\}$\;
$V^{(0)}, \ldots, V^{(N-1)} \leftarrow \emptyset$; $\mu_0 \leftarrow \mathbf{0}$\;
\For{$s \in \{A, B\}$}{
    $V^{(0)}[((1,s), (1,s), \textup{Upd}(\mu_0, 1, s))] \leftarrow 1$\;
}
\For{$\ell = 0$ \KwTo $N-2$}{
    \ForEach{$((u, s_u), (v, s_v), \mu) \in V^{(\ell)}$}{
        $\omega \leftarrow V^{(\ell)}[(u, s_u), (v, s_v), \mu]$\;
        \tcp{Extend}
        \For{$w \in \textup{Blk}(v) \setminus \{u\}$, $\mu[\textup{idx}(w)] \neq 2$}{
            $S_w \leftarrow \begin{cases}
                \{\mu[\textup{idx}(w)][1]\} & \text{if } \mu[\textup{idx}(w)] = (1,\cdot)\\
                \{A, B\} & \text{otherwise}
            \end{cases}$\;
            \For{$s_w \in S_w$}{
                $V^{(\ell+1)}[(u, s_u), (w, s_w), \textup{Upd}(\mu, w, s_w)] \mathrel{+}= \omega a_{v s_v, w s_w}$\;
            }
        }
        \tcp{Close}
        \If{$\exists k : \mu[k] = (1, s')$}{
            $(p, q) \leftarrow \min$ such interaction\;
            \For{$h \in \{p, q\} \setminus \{u\}$}{
                $V^{(\ell+1)}[(h, s'), (h, s'), \textup{Upd}(\mu, h, s')] \mathrel{+}= -\omega a_{v s_v, u s_u}$\;
            }
        }
    }
}
$\mathcal{S} \leftarrow \sum_{((u,s_u),(v,s_v),\mu) \in V^{(N-1)}} V^{(N-1)}[(u,s_u),(v,s_v),\mu] \cdot (-a_{v s_v, u s_u})$\;
\Return{$\mathcal{S}$}
\end{algorithm}
\end{figure*}

In Alg.~\ref{alg:cos_sublattice}, the $N=2n$ endpoints are grouped into interaction lines.  The mask $\mu$ records which interaction endpoints have been visited and, for singly occupied interactions, which sublattice label has been assigned.  The block constraint $\mathrm{Blk}(v)$ enforces the spin-block structure of the two-spin matrix.  The two transitions are the continuation and close-and-open moves described in the Results section, with edge labels $a_{v s_v,w s_w}$ carrying both spacetime and sublattice indices.

Algorithm~\ref{alg:cos_dedup} treats uniform flavors with a sublattice sum at each contact interaction. Its \texttt{Shift} routine canonicalizes equivalent endpoint slots while retaining the sublattice data needed by future factors. The visited set $R$ is stored before canonicalization.

Alg.~\ref{alg:cos_typed_heads} implements SB-CoS.  Each unpaired endpoint in $\mathcal{U}$ stores whether it is a normal or shift head and which flavor it carries.  Paired endpoints no longer need this payload, because their local color and shift information has already been fixed by the interaction partner.

The recurrences can be compiled into edges storing a source, destination, coefficient, and matrix or shift label. Evaluation then propagates numerical values over the stored edges for each Monte Carlo configuration. The all-subset readouts use the same representation with several output destinations.

\begin{figure*}
\begin{algorithm}[H]
\DontPrintSemicolon
\caption{Connected Sum for $\text{SU}(N_{f})$ Bipartite Model}
\label{alg:cos_dedup}
\KwIn{$n$ (interaction pairs)}
\KwOut{Sum $\mathcal{S}$ with $N_f$}
$N \leftarrow 2n$; $\mathcal{I} \leftarrow \{(i, i+n) : i \in [n]\}$\;
$V^{(0)}, \ldots, V^{(N-1)} \leftarrow \emptyset$; $\mu_0\leftarrow\mathbf 0$\;
\SetKwProg{Fn}{Func}{:}{}
\Fn{Shift$(u, v, R, \mu)$}{
    Cyclic shift mod $n$; map duplicates to partners\;
    \Return{shifted state}\;
}
\Fn{Add$(l, u, v, R, \mu, \omega)$}{
    $V^{(l)}[\textup{Shift}(u, v, R, \mu)] \mathrel{+}= \omega$\;
}
\For{$s \in \{A, B\}$}{
    Add$(0, (1,s), (1,s), \{1\}, \textup{Upd}(\mu_0, 1, s), 1)$\;
}
\For{$\ell = 0$ \KwTo $N-2$}{
    \ForEach{$((u, s_u), (v, s_v), R, \mu) \in V^{(\ell)}$}{
        $\omega \leftarrow V^{(\ell)}[(u, s_u), (v, s_v), R, \mu]$\;
        \tcp{Extend}
        \For{$w \in [N] \setminus R$, $\mu[\textup{idx}(w)] \neq 2$}{
            $S_w \leftarrow \begin{cases}
                \{\mu[\textup{idx}(w)][1]\} & \text{if } \mu[\textup{idx}(w)] = (1,\cdot)\\
                \{A, B\} & \text{otherwise}
            \end{cases}$\;
            \For{$s_w \in S_w$}{
                Add$(\ell\!+\!1, (u,s_u), (w,s_w), R \cup \{w\}, \textup{Upd}(\mu,w,s_w), \omega a_{v_n s_v, w_n s_w})$\;
            }
        }
        \tcp{Close}
        $\mathcal{P} \leftarrow \{q : (p,q) \in \mathcal{I}, p \in R, \mu[\textup{idx}(p,q)] = (1,\cdot)\}$\;
        \If{$\mathcal{P} \neq \emptyset$}{
            $h \leftarrow \min(\mathcal{P})$; $s_h \leftarrow \mu[\textup{idx}(h)][1]$\;
            Add$(\ell\!+\!1, (h,s_h), (h,s_h), R \cup \{h\}, \textup{Upd}(\mu,h,s_h), -\omega a_{v_n s_v, u_n s_u} N_f)$\;
        }
    }
}
$\mathcal{S} \leftarrow \sum_{((u,s_u),(v,s_v),R,\mu) \in V^{(N-1)}} V^{(N-1)}[(u,s_u),(v,s_v),R,\mu] \cdot (-a_{v_n s_v, u_n s_u} N_f)$\;
\Return{$\mathcal{S}$}
\end{algorithm}
\end{figure*}

\begin{figure*}
\begin{algorithm}[H]
\DontPrintSemicolon
\caption{Connected Sum for $\mathrm{SU}(N_f)$ Model with Flavor-Dependent Shifts and Edge Matrices}
\label{alg:cos_typed_heads}
\KwIn{$n$, matrices $A_\lambda\in\mathbb C^{n\times n}$, shifts $\sigma_\lambda$, $\lambda\in[N_f]$}
\KwOut{Physical connected sum $\mathcal S$}
$N\leftarrow2n$; $\bar v=1+((v-1)\bmod n)$; $P(v)=v+n$ for $v\le n$ and $v-n$ otherwise\;
$V^{(0)},\ldots,V^{(N-1)}\leftarrow$ zero-initialized sparse maps\;
\SetKwProg{Fn}{Function}{:}{}
\Fn{\textup{Shift}$((u,\tau,\lambda),v,R,\mathcal U)$}{
   $m_j\leftarrow|R\cap\{j,j+n\}|$ for $j\in[n]$\;
   $R_c\leftarrow\{j:m_j\ge1\}\cup\{j+n:m_j=2\}$\;
   $\mathcal U_c\leftarrow\{(\bar p,\tau_p,\lambda_p):(p,\tau_p,\lambda_p)\in\mathcal U\}$\;
   \Return{$((\bar u,\tau,\lambda),\bar v,R_c,\mathcal U_c)$}\;
}
\For{$\lambda \in [N_f]$, $\tau \in \{\text{normal}, \text{shift}\}$}{
   $q_0\leftarrow\textup{Shift}((1,\tau,\lambda),1,\{1\},\{(1,\tau,\lambda)\})$\;
   $V^{(0)}[q_0]\mathrel{+}=1$\;
}
\For{$i = 0$ \KwTo $N-2$}{
   \ForEach{$((u, \tau_u, \lambda_u), v, R, \mathcal{U}) \in V^{(i)}$}{
       $\omega \leftarrow V^{(i)}[(u, \tau_u, \lambda_u), v, R, \mathcal{U}]$\;
       \tcp{Rule 1: Continue cycle (only if normal head)}
       \If{$\tau_u = \text{normal}$}{
           \ForEach{$w \in [N] \setminus R$}{
               $R' \leftarrow R \cup \{w\}$; $\mathcal{U}' \leftarrow \mathcal{U}$\;
               \lIf{$P(w) \in \{x : (x, \cdot, \cdot) \in \mathcal{U}\}$}{Remove $(P(w), \cdot, \cdot)$ from $\mathcal{U}'$}
               \lElse{$\mathcal{U}' \leftarrow \mathcal{U}' \cup \{(w, \tau_u, \lambda_u)\}$}
               $q'\leftarrow\textup{Shift}((u,\tau_u,\lambda_u),w,R',\mathcal U')$\;
               $V^{(i+1)}[q']\mathrel{+}=\omega A_{\lambda_u}[\bar v,\bar w]$\;
           }
       }
       \tcp{Rule 2: Close cycle and start new}
       \uIf{$\tau_u=\mathrm{normal}$}{$\omega_c\leftarrow-\omega A_{\lambda_u}[\bar v,\bar u]$\;}
       \Else{$\omega_c\leftarrow-\omega\sigma_{\lambda_u}$\;}
       \If{$\mathcal U\neq\emptyset$}{
           $j\leftarrow\min\{\bar p:(p,\cdot,\cdot)\in\mathcal U\}$\;
           $(p_h,\tau_{p_h},\lambda_{p_h})\leftarrow$ the record in $\mathcal U$ with $\bar p_h=j$; $h\leftarrow P(p_h)$\;
           $R' \leftarrow R \cup \{h\}$; $\mathcal{U}' \leftarrow \mathcal{U} \setminus \{(p_h, \tau_{p_h}, \lambda_{p_h})\}$\;
           $\Lambda\leftarrow\{\lambda_{p_h}\}$ if $\tau_{p_h}=\mathrm{shift}$, otherwise $[N_f]$\;
           \For{$\lambda\in\Lambda$}{
               $q'\leftarrow\textup{Shift}((h,\mathrm{normal},\lambda),h,R',\mathcal U')$\;
               $V^{(i+1)}[q']\mathrel{+}=\omega_c$\;
           }
           $q'\leftarrow\textup{Shift}((h,\mathrm{shift},\lambda_{p_h}),h,R',\mathcal U')$\;
           $V^{(i+1)}[q']\mathrel{+}=\omega_c$\;
       }
   }
}
$\mathcal S\leftarrow0$\;
\ForEach{$q=((u,\tau_u,\lambda_u),v,R,\mathcal U)\in V^{(N-1)}$}{
   \uIf{$\tau_u=\mathrm{normal}$}{$\mathcal S\mathrel{-}=V^{(N-1)}[q]A_{\lambda_u}[\bar v,\bar u]$\;}
   \Else{$\mathcal S\mathrel{-}=V^{(N-1)}[q]\sigma_{\lambda_u}$\;}
}
\lIf{$n=1$}{$\mathcal S\mathrel{-}=\sum_{\lambda=1}^{N_f}\sigma_\lambda^2$}
\Return{$\mathcal{S}$}
\end{algorithm}
\end{figure*}

% ---------- In your document body ----------
\FloatBarrier
\subsection{Proof of Algorithm 3}
\label{sec:proof_algorithm3}

Let $n\ge2$ and fix the propagator matrices and shift-close weights at the interaction vertices. We construct the unique traversal of each connected typed contraction and show that canonicalization preserves its weight. The singleton correction is given at the end.

\paragraph{Diagrammatic objects.}

Fix a perturbative order $n$ and let
$V=[2n]=\{1,\ldots,2n\}$
be the set of fermionic endpoints. Interaction $i$ joins the two endpoints $i$ and $i+n$. The contact interactions therefore define the perfect matching
\begin{equation}
M=\bigl\{\{i,i+n\}:1\leq i\leq n\bigr\}.
\label{eq:alg3_matching}
\end{equation}
We denote the matching partner of $v$ by $P(v)$ and its interaction residue by
$\bar v=1+\bigl((v-1)\bmod n\bigr)$.

A Wick contraction of the $2n$ endpoints is represented by a permutation
$\pi\in S_{2n}$.
The edge $v\to\pi(v)$ is a fermion contraction, and the cycles of $\pi$ are its loops. For one flavor, introduce the $2n\times2n$ endpoint matrix $\mathcal A_{vw}=A[\bar v,\bar w]$ from the $n\times n$ residue matrix $A$. Its cycle-cover expansion is
\begin{equation}
\det\mathcal A
=
\sum_{\pi\in S_{2n}}
\operatorname{sgn}(\pi)
\prod_{v\in V}A[\bar v,\overline{\pi(v)}].
\end{equation}

Flavor conservation assigns one label to each normal cycle $C$,
$\lambda(C)\in[N_f]$.
Normal cycles have independent color labels, including equal-color assignments in the formal contact-interaction sum.

The counterterm can replace the propagator factor associated with a one-cycle. We call such a replaced one-cycle a \emph{shift cycle}. Let
$\mathcal{Z} \subseteq \bigl\{\{v\}: \pi(v)=v\bigr\}$
be the set of shift cycles. Cycles outside $\mathcal{Z}$ are called normal cycles.

A shift cycle inherits the color of the normal cycle containing its matching partner. For $\{v\}\in\mathcal{Z}$,
\begin{equation}
\lambda_{\mathrm{sh}}(v)
=
\lambda\bigl(C_\pi(P(v))\bigr),
\label{eq:alg3_shift_flavor}
\end{equation}
where $C_\pi(w)$ denotes the permutation cycle containing $w$.

For $n\geq2$, both endpoints of a matching pair cannot be shifts in a connected diagram. Their only permutation edges would be
$v\longrightarrow v, \qquad P(v)\longrightarrow P(v)$,
and the matching edge $\{v,P(v)\}$ would leave the pair isolated from all other endpoints.

The undirected graph associated with $\pi$ is
\begin{equation}
G(\pi)
=
\left(
V,\,
\bigl\{\{v,\pi(v)\}:v\in V\bigr\}\cup M
\right).
\label{eq:alg3_augmented_graph}
\end{equation}
The contraction is connected if and only if $G(\pi)$ is connected.

A typed contraction is a triple
$\Omega=(\pi,\mathcal{Z},\lambda)$,
consisting of a permutation, a set of shifted one-cycles, and one flavor for each normal cycle. Its algebraic weight is
\begin{align}
W(\Omega)
={}&
\operatorname{sgn}(\pi)
\prod_{\substack{C\in\operatorname{Cyc}(\pi)\\C\notin\mathcal{Z}}}
\prod_{v\in C}
A_{\lambda(C)}
[\bar v,\overline{\pi(v)}]
\nonumber\\
&\times
\prod_{\{v\}\in\mathcal{Z}}
\sigma_{\lambda_{\mathrm{sh}}(v)} .
\label{eq:alg3_target_weight}
\end{align}
Here $\sigma_\lambda$ is the shift-close weight for flavor $\lambda$. Let $\mathfrak{C}_n$ be the set of typed contractions for which $G(\pi)$ is connected. The desired connected sum is
\begin{equation}
\mathcal{S}_n
=
\sum_{\Omega\in\mathfrak{C}_n}
W(\Omega).
\label{eq:alg3_target_sum}
\end{equation}

\paragraph{Unquotiented typed-head recurrence.}

With absolute endpoint labels, the state is
\begin{equation}
q=
\bigl((u,\tau_u,\lambda_u),v,R,\mathcal{U}\bigr).
\label{eq:alg3_raw_state}
\end{equation}
Here $R\subseteq V$ is the visited set, $u\in R$ the head, and $v\in R$ the current endpoint. The head has type $\tau_u\in\{\mathrm{normal},\mathrm{shift}\}$ and flavor label $\lambda_u\in[N_f]$. For a shift head, this label specifies its partner flavor. The unpaired interactions carry the records
\begin{equation}
\mathcal{U}
=
\left\{
(p,\tau_p,\lambda_p):
p\in R,\ P(p)\notin R
\right\}.
\label{eq:alg3_unpaired_set}
\end{equation}
Each singly occupied interaction contributes one record.

The recurrence starts at endpoint $1$. For every flavor $\lambda$, it creates one normal initial state and one shift initial state:
\begin{align}
q_{\mathrm{N},\lambda}^{(0)}
&=
\bigl((1,\mathrm{normal},\lambda),1,\{1\},
      \{(1,\mathrm{normal},\lambda)\}\bigr),\\
q_{\mathrm{S},\lambda}^{(0)}
&=
\bigl((1,\mathrm{shift},\lambda),1,\{1\},
      \{(1,\mathrm{shift},\lambda)\}\bigr).
\end{align}
The initial value of every such state is one.

If the current head is normal, Rule~1 may append any unvisited endpoint $w$. The transition factor is
\begin{equation}
A_{\lambda_u}[\bar v,\bar w].
\label{eq:alg3_continue_factor}
\end{equation}
If $P(w)$ is unvisited, insert
$(w,\mathrm{normal},\lambda_u)$
into $\mathcal{U}$. Otherwise the interaction is completed and its partner record is removed.

A shift head represents one replaced one-cycle and closes immediately.

Rule~2 closes the current cycle. The closing factor is
$-\, A_{\lambda_u}[\bar v,\bar u]$
for a normal head and
$-\sigma_{\lambda_u}$
for a shift head.

After a cycle has been closed, the visited set is a union of complete permutation cycles. If every endpoint has been visited, this closing factor is sent to the final output. Otherwise the algorithm must open a new cycle.

Let
\begin{equation}
\partial_M R
=
\{h\in V\setminus R:P(h)\in R\}
\label{eq:alg3_frontier}
\end{equation}
be the matching frontier. Its elements are the unvisited partners of the endpoints represented in $\mathcal{U}$. Order the interactions by their residues and choose the smallest one in this frontier; its unvisited endpoint is $h$. This ordering is unchanged by exchanging the two slots of an interaction. Let
$p=P(h)$
be the already visited endpoint and let
$(p,\tau_p,\lambda_p)\in\mathcal{U}$
be its stored payload. Once $h$ is visited, this interaction becomes fully occupied and the record for $p$ is removed.

The normal successors are
\begin{equation}
\bigl((h,\mathrm{normal},\lambda),h,
      R\cup\{h\},\mathcal{U}'\bigr),
\qquad
\lambda\in\Lambda_p,
\label{eq:alg3_normal_successors}
\end{equation}
where
\begin{equation}
\Lambda_p=
\begin{cases}
[N_f],&\tau_p=\mathrm{normal},\\
\{\lambda_p\},&\tau_p=\mathrm{shift}.
\end{cases}
\label{eq:alg3_normal_label_rule}
\end{equation}
There is also one shift successor,
\begin{equation}
\bigl((h,\mathrm{shift},\lambda_p),h,
      R\cup\{h\},\mathcal{U}'\bigr).
\label{eq:alg3_shift_successor}
\end{equation}
The type and color of the new head are part of the successor keys in Eqs.~\eqref{eq:alg3_normal_successors} and \eqref{eq:alg3_shift_successor}.

\begin{lemma}[Typed-frontier invariant]
\label{lem:alg3_typed_frontier_invariant}
At every layer of the recurrence, $\mathcal{U}$ contains exactly the visited endpoints whose matching partners are unvisited. Its payload is sufficient to determine all future flavor constraints.
\end{lemma}

\begin{proof}
Initially, $\mathcal U$ contains the record for endpoint $1$. A continuation to an untouched interaction inserts a record; continuation to its partner removes it. Rule~2 always consumes the unvisited partner of a stored endpoint and removes that record. These updates establish the membership invariant by induction.

A normal record stores its cycle color, and a shift record stores the color required at its normal partner. An unmatched shift record can arise only from initialization: that head closes without continuation and immediately opens its own matching partner. Every later shift head is opened at a partner that has already been visited, so it creates no unmatched record. A normal continuation thus never encounters a pending shift-color constraint; that constraint has already been imposed when its head opened. After an interaction is completed, its frontier payload can be discarded.
\end{proof}

\paragraph{Existence of the canonical cycle ordering.}

We next show that the frontier required by Rule~2 cannot become empty in a connected contraction before every cycle has been processed.

\begin{lemma}[Nonempty matching frontier]
\label{lem:alg3_nonempty_frontier}
Let $\pi\in S_{2n}$ be such that $G(\pi)$ is connected. If
$\varnothing\neq R\subsetneq V$
is a union of complete cycles of $\pi$, then
$\partial_M R\neq\varnothing$.
\end{lemma}

\begin{proof}
Because $R$ is a union of complete permutation cycles,
$v\in R\quad\Longrightarrow\quad\pi(v)\in R$.
Therefore no permutation edge crosses the cut $(R,V\setminus R)$.

Suppose that $\partial_M R$ were empty. Then no matching edge in $M$ would cross the same cut. Neither the permutation edges nor the matching edges would connect $R$ to $V\setminus R$, and $G(\pi)$ would be disconnected. This contradicts the assumption.
\end{proof}

For a connected permutation, let $C_1$ be the cycle containing endpoint $1$ and set
$R_1=C_1$.
If $R_t\neq V$, Lemma~\ref{lem:alg3_nonempty_frontier} guarantees at least one crossing matching edge. Choose the first such interaction in the fixed ordering, let $h_{t+1}$ be its unvisited endpoint, and let
$C_{t+1}=C_\pi(h_{t+1})$.
Then define
$R_{t+1}=R_t\cup C_{t+1}$.

Since $h_{t+1}\notin R_t$, the selected cycle is new and $R_{t+1}$ strictly enlarges $R_t$. Repetition terminates with the unique cycle ordering
\begin{equation}
C_1,C_2,\ldots,C_k
\label{eq:alg3_canonical_cycle_order}
\end{equation}
of all cycles of $\pi$. This is the canonical cycle order used by Algorithm~3.

\begin{lemma}[Every complete algorithmic path is connected]
\label{lem:alg3_path_connected}
Every complete path generated by the unquotiented recurrence defines a connected augmented graph.
\end{lemma}

\begin{proof}
The first completed cycle forms the initial connected component. Whenever Rule~2 opens a new cycle at $h$, the endpoint $P(h)$ has already been visited. The matching edge
$\{P(h),h\}\in M$
therefore joins the new cycle to the previously processed union.

By induction, the union of all completed cycles and the matching edges used to open them is connected. A complete path visits all $2n$ endpoints, so its full augmented graph is connected.
\end{proof}

\paragraph{Bijection between paths and connected contractions.}

\begin{lemma}[Surjectivity]
\label{lem:alg3_surjectivity}
Every connected typed contraction
$\Omega\in\mathfrak{C}_n$
is generated by at least one complete path of the unquotiented recurrence.
\end{lemma}

\begin{proof}
Let
$C_1,\ldots,C_k$
be the canonical cycle order of the permutation in $\Omega$.

If $C_1$ is normal, initialize a normal head at endpoint $1$ with flavor $\lambda(C_1)$. If $C_1=\{1\}$ is a shift cycle, initialize a shift head with the flavor of the normal cycle containing $P(1)$. All flavors occur among the initial states, so the required initial state exists.

Consider a normal cycle
$C_t=(v_0,v_1,\ldots,v_{m-1})$,
written in its directed permutation order:
$\pi(v_j)=v_{j+1}, \qquad \pi(v_{m-1})=v_0$.
Its canonical head is $v_0$. Apply Rule~1 successively to
$v_1,\ldots,v_{m-1}$.
All these vertices are unvisited because permutation cycles are disjoint. The flavor remains $\lambda(C_t)$ along the whole cycle. Rule~2 then closes the edge $v_{m-1}\rightarrow v_0$.

If $C_t$ is a shift cycle, it contains only its head and is closed immediately with no Rule~1 transition.

Suppose that the next canonical head is $h$ and that
$p=P(h)$
belongs to an already processed cycle. There are three connected possibilities.

If both the cycle containing $p$ and the new cycle are normal, the two normal flavors are independent. Equation~\eqref{eq:alg3_normal_label_rule} allows all $\lambda\in[N_f]$ and therefore includes the required flavor of the new cycle.

If $p$ belongs to a shift cycle, the new cycle must be normal and must have flavor $\lambda_p$. Equation~\eqref{eq:alg3_normal_label_rule} enforces exactly this choice.

If the new cycle is a shift cycle, then $p$ belongs to a normal cycle and Eq.~\eqref{eq:alg3_shift_successor} assigns the shift head the flavor $\lambda_p$ of that normal cycle.

Thus every canonical cycle can be generated with its prescribed type and flavor. Repeating the construction produces a complete path representing $\Omega$.
\end{proof}

\begin{lemma}[Injectivity]
\label{lem:alg3_injectivity}
Two different complete paths of the unquotiented recurrence cannot represent the same connected typed contraction.
\end{lemma}

\begin{proof}
A path specifies the permutation cycles, the shifted one-cycles, and the normal-cycle colors. Conversely, the contraction fixes the cycle containing endpoint $1$ and every continuation along that cycle. After closure, the ordered matching frontier fixes the next head. Applying the same rule to each cycle fixes the full path. Distinct paths must differ in the permutation, a head type, or a normal-cycle color.
\end{proof}

Combining Lemmas~\ref{lem:alg3_surjectivity} and \ref{lem:alg3_injectivity} gives a bijection
\begin{equation}
\left\{
\text{complete unquotiented paths}
\right\}
\longleftrightarrow
\mathfrak{C}_n.
\label{eq:alg3_path_bijection}
\end{equation}

\paragraph{Equality of path and diagrammatic weights.}

\begin{lemma}[Weight preservation]
\label{lem:alg3_weight_preservation}
Under the bijection in Eq.~\eqref{eq:alg3_path_bijection}, the product of all transition factors along a path is equal to the diagrammatic weight in Eq.~\eqref{eq:alg3_target_weight}.
\end{lemma}

\begin{proof}
Consider a normal cycle
$C=(v_0,v_1,\ldots,v_{m-1})$
with flavor $\lambda(C)$. Its Rule~1 transitions contribute
\begin{equation}
\prod_{j=0}^{m-2}
A_{\lambda(C)}[\bar v_j,\bar v_{j+1}],
\end{equation}
and Rule~2 contributes the closing factor
$- A_{\lambda(C)}[\bar v_{m-1},\bar v_0]$.
The complete normal-cycle contribution is
\begin{equation}
-
\prod_{v\in C}
A_{\lambda(C)}
[\bar v,\overline{\pi(v)}].
\label{eq:alg3_normal_cycle_weight}
\end{equation}

A shift cycle $\{v\}$ contributes
\begin{equation}
-\sigma_{\lambda_{\mathrm{sh}}(v)}.
\label{eq:alg3_shift_cycle_weight}
\end{equation}
The inherited flavor is correct by Eqs.~\eqref{eq:alg3_normal_label_rule} and \eqref{eq:alg3_shift_successor}.

Let $k$ be the total number of permutation cycles, including shifted one-cycles. Equations~\eqref{eq:alg3_normal_cycle_weight} and \eqref{eq:alg3_shift_cycle_weight} provide one minus sign per cycle, so the algorithmic sign is
$(-1)^k$.
The sign of a permutation of $2n$ objects with $k$ cycles is
$\operatorname{sgn}(\pi) = (-1)^{2n-k}$.
Since $2n$ is even,
\begin{equation}
(-1)^{2n-k}=(-1)^k.
\label{eq:alg3_sign_equality}
\end{equation}
This is the determinant sign.

Multiplying the cycle weights gives Eq.~\eqref{eq:alg3_target_weight}. The normal-color branches perform the flavor sum, so the typed recurrence needs no additional loop factor $N_f$.
\end{proof}

\paragraph{Contact-interaction quotient.}

The implementation canonicalizes the absolute endpoints of each contact interaction. Instead of storing which of the two endpoints has been visited, it stores only
$m_r\in\{0,1,2\}$,
the number of visited endpoints of interaction $r$. If $m_r=1$, it also stores the type and flavor payload of the visited endpoint.

The canonical state can therefore be written as
\begin{equation}
\widehat q=
\bigl(
\bar u,\bar v,\tau_u,\lambda_u,
\mathbf m,\boldsymbol{\eta}
\bigr),
\label{eq:alg3_canonical_state}
\end{equation}
where
\begin{equation}
\eta_r=
\begin{cases}
(\tau_r,\lambda_r),&m_r=1,\\
\varnothing,&m_r\in\{0,2\}.
\end{cases}
\end{equation}
The operation denoted by $\operatorname{Shift}$ in Alg.~\ref{alg:cos_typed_heads} maps an absolute state to this canonical representation.

\begin{lemma}[Endpoint multiplicity]
\label{lem:alg3_endpoint_multiplicity}
Suppose a normal cycle continues into interaction $r$. The number of absolute endpoints producing the same canonical transition is
$2-m_r$.
\end{lemma}

\begin{proof}
For $m_r=0$, the two unvisited endpoints have the same residue, factor, and canonical child. For $m_r=1$, only the matching partner is available. For $m_r=2$, neither endpoint is available. The respective multiplicities are two, one, and zero.
\end{proof}

\begin{lemma}[Quotient preservation]
\label{lem:alg3_quotient_preservation}
When every canonical continuation edge is assigned the multiplicity $2-m_r$, the value propagated by the canonicalized recurrence is exactly the sum of the values propagated by all corresponding absolute-state paths.
\end{lemma}

\begin{proof}
Let $Q$ map an absolute partial path to its canonical state. We prove by induction over the layer $\ell$ that
\begin{equation}
\widehat V^{(\ell)}(\widehat q)
=
\sum_{\substack{p\ \mathrm{absolute\ partial\ path}\\
                 Q(p)=\widehat q}}
W_\ell(p),
\label{eq:alg3_quotient_invariant}
\end{equation}
where $W_\ell(p)$ is the product of the transition factors accumulated along $p$.

At layer zero, the first endpoint is fixed to its canonical representative, and Eq.~\eqref{eq:alg3_quotient_invariant} holds with coefficient one.

Assume that it holds at layer $\ell$. For a Rule~1 transition, Lemma~\ref{lem:alg3_endpoint_multiplicity} gives the exact number of absolute endpoint choices represented by each canonical edge. These absolute choices have identical matrix factors because
$\mathcal A_{\lambda;v,w} = A_\lambda[\bar v,\bar w]$.
They also have the same canonical child and the same typed unpaired payload.

For Rule~2, the selected interaction is singly occupied. Its unvisited matching partner is unique, so its multiplicity is one. The type and flavor of the new head remain explicit and are not identified by canonicalization.

Each group of absolute successors contributes to its canonical child with the stated multiplicity. This proves Eq.~\eqref{eq:alg3_quotient_invariant} at layer $\ell+1$ and, by induction, at the final readout.
\end{proof}

\begin{theorem}[Correctness of Algorithm 3]
\label{thm:alg3_correctness}
For every $n\geq2$, every number of flavors $N_f$, arbitrary flavor-dependent matrices
$A_1,\ldots,A_{N_f}$,
and arbitrary flavor-dependent shift-close weights
$\sigma_1,\ldots,\sigma_{N_f}$,
Alg.~\ref{alg:cos_typed_heads} returns the connected typed contraction sum
\begin{equation}
\mathcal{S}
=
\sum_{\Omega\in\mathfrak{C}_n}
W(\Omega).
\end{equation}
\end{theorem}

\begin{proof}
Lemmas~\ref{lem:alg3_path_connected}--\ref{lem:alg3_injectivity} identify complete paths with connected typed contractions. Lemma~\ref{lem:alg3_weight_preservation} equates their weights, and Lemma~\ref{lem:alg3_quotient_preservation} preserves their sum under canonicalization. The final output is $\mathcal S_n$.
\end{proof}
At $n=1$, the raw recurrence also generates two matched shift one-cycles of the same color, contributing $\sum_\lambda\sigma_\lambda^2$. The last line of Algorithm~\ref{alg:cos_typed_heads} removes this field-independent term, leaving
\begin{equation}
\mathcal S_1=2\sum_{a<b}A^{(a)}_{11}A^{(b)}_{11}
+2\sum_a\sigma_a A^{(a)}_{11}.
\label{eq:physical_singleton}
\end{equation}
For $n\ge2$, two matched shift one-cycles form an isolated component and cannot reach a connected full-order output. Figure~\ref{fig:example_diagrams} shows representative shifted contractions.
\begin{figure}[htbp]
    \centering
    \begin{tikzpicture}[
        wiggle/.style={
            black, 
            line width=1pt,
            decorate,
            decoration={snake, amplitude=1.2mm, segment length=4mm, pre length=1mm, post length=1mm}
        },
        rline/.style={myred, line width=1.5pt},
        bline/.style={myblue, line width=1.5pt},
        gline/.style={mygreen, line width=1.5pt}
    ]

    % Helper macro to draw perfect crosses easily
    \newcommand{\drawcross}[3][]{
        \draw[#1] (#2) ++(-#3,#3) -- ++({2*#3},{-2*#3});
        \draw[#1] (#2) ++(-#3,-#3) -- ++({2*#3},{2*#3});
    }

    % ==========================================
    % 布局控制参数 (在这里调整距离和标签位置)
    % ==========================================
    \def\xshift{4}  % 左右子图的间距
    \def\yshift{-6} % 上下子图的间距
    \def\labely{-4.2} % (a)(b)(c)(d) 标签的统一纵坐标 (确保绝对水平对齐)

    % =========================
    % TOP-LEFT PANEL (a)
    % =========================
    \begin{scope}[shift={(0,0)}]
        % Wavy lines
        \draw[wiggle] (0,0) -- (0,-3);
        \draw[wiggle] (2,0) -- (2,-3);

        % Red Leaf
        \draw[rline] (0,-3) to[bend left=20] (2,0);
        \draw[rline] (0,-3) to[bend right=20] (2,0);

        % Red Crosses
        \drawcross[rline]{0,0}{0.2}
        \drawcross[rline]{2,-3}{0.2}
        
        % Subfigure label (居中在 x=1)
        \node at (1, \labely) {(a)};
    \end{scope}

    % =========================
    % TOP-RIGHT PANEL (b)
    % =========================
    \begin{scope}[shift={(\xshift,0)}]
        % Wavy lines
        \draw[wiggle] (0,0) -- (0,-3);
        \draw[wiggle] (1.5,0) -- (1.5,-3);
        \draw[wiggle] (3,0) -- (3,-3);

        % Blue Arcs
        \draw[bline] (0,0) to[bend left=45] (1.5,0);
        \draw[bline] (1.5,0) to[bend left=45] (3,0);
        \draw[bline] (0,0) to[bend left=60] (3,0);

        % Blue Cross
        \drawcross[bline]{0,-3}{0.2}

        % Blue Circles
        \draw[bline, fill=white] (1.5,-3.2) circle (0.3);
        \draw[bline, fill=white] (3,-3.2) circle (0.3);
        
        % Subfigure label (居中在 x=1.5)
        \node at (1.5, \labely) {(b)};
    \end{scope}

    % =========================
    % BOTTOM-LEFT PANEL (c)
    % =========================
    \begin{scope}[shift={(0,\yshift)}]
        % Wavy lines (Middle one is shorter to match sketch)
        \draw[wiggle] (0,0) -- (0,-3);
        \draw[wiggle] (1.5,0) -- (1.5,-2);
        \draw[wiggle] (3,0) -- (3,-3);

        % Blue Top Arc
        \draw[bline] (0,0) to[bend left=45] (3,0);
        
        % Blue hanging V-swoop 
        \draw[bline] (0,0) to[bend right=15] (1.5,-2);
        \draw[bline] (3,0) to[bend left=15]  (1.5,-2);

        % Green Leaf (Bottom)
        \draw[gline] (0,-3) to[bend left=20] (3,-3);
        \draw[gline] (0,-3) to[bend right=20] (3,-3);

        % Blue Cross
        \drawcross[bline]{1.5, 0}{0.2}
        
        % Subfigure label (居中在 x=1.5)
        \node at (1.5, \labely) {(c)};
    \end{scope}

    % =========================
    % BOTTOM-RIGHT PANEL (d)
    % =========================
    \begin{scope}[shift={(\xshift,\yshift)}]
        % Three Wavy lines
        \draw[wiggle] (0,0) -- (0,-3);
        \draw[wiggle] (1.5,0) -- (1.5,-3);
        \draw[wiggle] (3,0) -- (3,-3);

        % Red Cross (Top left)
        \drawcross[rline]{0,0}{0.2}

        % Red Leaf (Bottom, connecting line 1 and 2)
        \draw[rline] (0,-3) to[bend left=30] (1.5,-3);
        \draw[rline] (0,-3) to[bend right=30] (1.5,-3);

        % Green Loop (Connecting line 2 and 3)
        \draw[gline] (1.5,0) to[bend left=45] (3,0);       % Top arc
        \draw[gline] (3,0) to[bend left=60] (3,-3);        % Right outward arc
        \draw[gline] (3,-3) to[bend right=15] (1.5,0);   % Bottom arc looping back to middle line
        
        % Subfigure label (居中在 x=1.5)
        \node at (1.5, \labely) {(d)};
    \end{scope}
    \end{tikzpicture}
    
    \caption{Shifted contractions in Algorithm~\ref{alg:cos_typed_heads}. Each normal fermion loop carries one color. A cross replaces a one-cycle by a counterterm whose color is inherited from the normal loop at its interaction partner.}
    \label{fig:example_diagrams}
\end{figure}

\begin{acknowledgments}
The author acknowledges computational resources and support from the Imperial College Research Computing Service (\url{http://doi.org/10.14469/hpc/2232}) and support from the Imperial College President's Scholarship.
\end{acknowledgments}

\FloatBarrier

\bibliographystyle{apsrev4-2}
% Use the precompiled bibliography; no BibTeX run is required for this package.
\providecommand{\noopsort}[1]{}\providecommand{\singleletter}[1]{#1}%

\end{document}